\documentclass[a4paper,12pt,english]{article}

\usepackage{amsfonts,bm,amssymb,euscript,array,babel,cite}
\usepackage{amsmath,amsthm}
\usepackage[dvips]{epsfig}
\usepackage{slashed}



\newcommand{\be}{\begin{equation}} \newcommand{\ee}{\end{equation}}
\newcommand{\beq}{\begin{equation}} \newcommand{\eeq}{\end{equation}}
\newcommand{\beqa}{\begin{eqnarray}}
\newcommand{\eeqa}{\end{eqnarray}} \newcommand{\eq}[1]{(\ref{#1})}
\def\nn{\nonumber} \def\bea{\begin{eqnarray}} \def\eea{\end{eqnarray}}
\def\obar{\overline}

\newcommand{\barr}{\begin{array}}
\newcommand{\earr}{\end{array}}

\def\a{\alpha}  \def\b{\beta}
 \def\g{\gamma} \def\G{\Gamma}
  \def\D{\Delta}

  \def\la{\lambda} 
    
\def\s{\sigma}

  \def\cC{{\cal C}}    
\def\cH{{\cal H}}      \def\cN{{\cal N}}
    \def\cS{{\cal S}}

\def\R{{\mathbb R}} \def\C{{\mathbb C}} \def\N{{\mathbb N}}
 \def\one{\mbox{1 \kern-.59em {\rm l}}}

\def\bit{\begin{itemize}} \def\eit{\end{itemize}} \def\Tr{\mbox{Tr}}

\def\({\left(} \def\){\right)} \def\tens{\otimes}

 \allowdisplaybreaks[3]

\begin{document}

\renewcommand{\title}[1]{\vspace{10mm}\noindent{\Large{\bf
#1}}\vspace{8mm}} \newcommand{\authors}[1]{\noindent{\large
#1}\vspace{5mm}} \newcommand{\address}[1]{{\itshape #1\vspace{2mm}}}

\begin{titlepage}
\begin{flushright}
UWThPh-2008-09
\end{flushright}

\begin{center}

\title{ \Large On the fermion spectrum of spontaneously generated \\[1ex]
fuzzy extra dimensions with fluxes}

\vskip 3mm

\authors{Athanasios {Chatzistavrakidis${}^{1,2}$},
Harold {Steinacker${}^3$},  \\[1ex]
 George {Zoupanos${}^2$}}

\vskip 3mm

\address{${}^1$ {\it Institute of Nuclear Physics,
NCSR  Demokritos,\\
GR-15310 Athens, Greece}\\[1ex] ${}^2$ Physics Department, National
Technical University,\\ Zografou Campus, GR-15780 Athens, Greece\\[1ex]
${}^3$Department of Physics, University of Vienna,\\
Boltzmanngasse 5, A-1090 Wien, Austria\\[3ex] E-mails:
{cthan@mail.ntua.gr,\\ harold.steinacker@univie.ac.at,\\ George.Zoupanos@cern.ch}}

\vskip 1.4cm

\textbf{Abstract}

\vskip 3mm

\begin{minipage}{14cm}%

We consider certain vacua of four-dimensional $SU(N)$ 
gauge theory with the same field content as the ${\cal N}=4$ 
supersymmetric Yang-Mills theory, resulting from potentials 
which break the $\cN=4$ supersymmetry as well as its global $SO(6)$ 
symmetry down to $SO(3) \times SO(3)$.
We show that the theory behaves at intermediate scales as
Yang-Mills theory on $M^4 \times S_L^2 \times S_R^2$,
where the extra dimensions are fuzzy spheres with magnetic fluxes.
We determine in particular the structure of the zero modes due to the
fluxes, which leads to low-energy mirror models.

%

\end{minipage}

\end{center}

\end{titlepage}

\tableofcontents


\section{Introduction}

The unification of the fundamental interactions has always been one of the main goals of theoretical physics. Several approaches have been employed in order to achieve this goal, one of the most exciting ones being the proposal that extra dimensions may exist in nature. The realization that superstring theories can  be consistently defined only in ten dimensions has led to an intensive study of possible compactifications of these theories with the hope that phenomenologically viable four-dimensional vacua can be revealed upon dimensional reduction. Recently a surprising new approach was proposed in \cite{Arkani-Hamed:2001ca}, where the above procedure was inversed.
In particular it was found that extra dimensions can arise dynamically
within a 
four-dimensional renormalizable and asymptotically-free gauge theory, as an effective description valid up to some energy scale. 
This has become known under the name of deconstruction.

A simple realization of the idea of a spontaneous
generation of  extra dimensions was given in
\cite{Aschieri:2006uw} and \cite{Steinacker:2007ay}, inspired 
by an earlier work \cite{Aschieri:2003vy} where the general ideas of \cite{Forgacs:1979zs,Kapetanakis:1992hf} were followed. In particular, it was shown in \cite{Aschieri:2006uw} that starting with the $SU(N)$ Yang-Mills theory on the Minkowski spacetime $M^4$ for some generic
(large) $N \in \N$ coupled with three scalars and adding the most 
general renormalizable $SO(3)$-invariant potential, 
an extra-dimensional fuzzy sphere is formed
via the Higgs effect.  The unbroken gauge group is generically
$SU(n_1) \times SU(n_2)
\times U(1)$, or possibly $SU(n)$. In \cite{Steinacker:2007ay} fermions were added in the previous model and their effective description from both the 6D and 4D 
point of view was worked out. The most interesting feature is that the extra-dimensional sphere then automatically  
carries a magnetic flux, which couples to 
the fermions transforming
in the bifundamental of $SU(n_1) \times SU(n_2)$. In view of this feature, the possibility to obtain chiral 4D models was studied. The outcome of this analysis was that only a picture of mirror fermions can be achieved in four dimensions\footnote{For a discussion on phenomenological aspects of such models see e.g. \cite{Maalampi:1988va}.}.

In the present work we explore the dynamical generation of a product
of two fuzzy spheres. 
In particular, we start with the $SU(N)$ Yang-Mills theory in four
dimensions, coupled to six scalars and four Majorana spinors, 
i.e. with the particle spectrum of the $\cN=4$ supersymmetric
Yang-Mills theory (SYM). Adding an explicit $R$-symmetry-breaking
potential, thus breaking the $\cN=4$ supersymmetry partially or completely,
we reveal stable $M^4\times S_L^2\times S^2_R$ vacua. In the most interesting case we include magnetic fluxes on the extra-dimensional fuzzy spheres and study the fermion spectrum, in particular the zero modes of the Dirac operator. The outcome of our analysis is that we obtain again a mirror model in low energies. However, in the present case we are able to single out an action which leads to exact separation of the two chiral mirror sectors which arise in the model.

One of the reasons for considering the $\cN=4$ SYM is that 
it is closely related to the IIB matrix model \cite{Ishibashi:1996xs}, which is 
a candidate for a quantum theory of fundamental interactions 
including gravity. The mechanism for gravity in that model was
clarified recently in \cite{Steinacker:2008ri}, where space-time is described by a 
noncommutative space. On $\R^4_\theta$, the matrix model is 
nothing but noncommutative $\cN=4$ SYM. 
Since extra dimensions can be realized in terms of 
deconstruction starting from a four-dimensional gauge theory, 
it is natural to look for stringy constructions such as
branes in extra dimensions (see e.g. \cite{Blumenhagen:2006ci} and references therein),
with the aim to recover the standard model or some extension of it.
In the present work we take some new steps in that direction, 
starting from commutative $\cN=4$ SYM. We 
recover products of branes (in the 
form of fuzzy spheres) with fluxes, and 
matter fields realized as bi-modules connecting the branes. 
While some aspects of the brane constructions are still missing, 
the present approach also offers advantages. In particular, 
our results are obtained within a renormalizable 
four-dimensional gauge theory,
and the vacua are at least local minima
without flat directions and unstable moduli. 

There has been considerable work on 
fuzzy geometries arising in matrix models,
see e.g. \cite{Azuma:2004qe,Aoki:2004sd,Grosse:2004wm} and references therein.
In particular, the case of 
fuzzy $S^2\times S^2$ has been studied 
in  \cite{Imai:2003ja}, and its gauge theory in \cite{Behr:2005wp}. 
The novel aspect of
the present paper is to take into account fluxes on these 
fuzzy spaces, and to study 
the fermionic zero modes due to these fluxes in extra
dimensions. For further literature on 
fuzzy spaces with fluxes see e.g.
\cite{Carow-Watamura:1998jn,Balachandran:2003ay,Steinacker:2003sd,Steinacker:2007iq,Aoki:2004sd,Aoki:2006wv,Dolan:2002ck,Aoki:2009cv},
and related work on the reduction of Yang-Mills models on $M^4 \times S^2$ 
in the presence of fluxes was given in 
\cite{Harland:2009kg,Dolan:2009ie}.

The outline of this article is as follows. In section 2 the action for the four-dimensional model that we consider is presented and the dynamical generation of the product of two fuzzy spheres is discussed. Moreover, the operators on $S^2_L\times S^2_R$ which are relevant in our analysis are defined, including the Dirac operator. In section 3 we add magnetic fluxes on the extra-dimensional fuzzy spheres and treat in detail the zero modes of the Dirac operator in order to discuss the chirality issue. It is shown that a mirror model is obtained and the separation of the two exactly chiral mirror sectors is discussed. In section 4 we present our conclusions and discuss the prospects of further work on the subject. Finally, in appendix A our Clifford algebra conventions are collected, while in appendix B the Dirac operator and its eigenmodes on the fuzzy sphere are presented.

\section{Yang-Mills gauge theory
and spontaneously generated fuzzy
$S^2\times S^2$}

\subsection{The action}

We consider the $SU(N)$ Yang-Mills gauge theory in four-dimensional 
Minkowski spacetime, coupled to six scalars
$\Phi_a = \Phi_a^\dagger$ ($a=1\dots 6$) and
four Majorana spinors $\chi_p$ ($p=1\dots 4$) in the adjoint representation 
of the $SU(N)$.
Moreover, we assume that the $\Phi_a$ transform in the vector representation 
of a global $SU(4) \cong SO(6)$ group and
the $\chi_p$ in the fundamental of the $SU(4)$. The above particle spectrum 
coincides with the spectrum of the ${\cal N}=4$ supersymmetric Yang-Mills 
theory (SYM) \cite{Brink:1976bc}, where the global $SU(4)$ is the $R$-symmetry of the theory.
The corresponding action, which is a modification of the ${\cal N}=4$ 
SYM theory, is given by
\bea
S_{YM} &=& \int d^4 x\Biggl[\,\Tr\(-\frac 1{4} F_{\mu\nu}F^{\mu\nu}
+ \frac 12 \sum\limits_{a=1}^6\,D^\mu\Phi_a D_\mu \Phi_a
- V(\Phi)\)  \nn\\
&& \quad\,+ \,\frac 12 \Tr \Big(i\bar\chi_p \slashed D \chi_p
  + g_4 (\Delta_R^a)_{pq}\,\bar\chi_p R [\Phi_a,\chi_q]
  - g_4 (\Delta_L^a)_{pq}\,\bar\chi_p L [\Phi_a,\chi_q] \Big)\Biggl],
\label{YM-action-4D}
\eea where the potential has the form \beq\label{4dpot}
 V(\Phi) = V_{{\cal N}=4}(\Phi)+ V_{break}(\Phi). \eeq In (\ref{4dpot}) 
the first term corresponds to the potential of the ${\cal N}=4$ SYM 
theory, which is explicitly given by
 \be\label{n4pot} V_{{\cal N}=4}(\Phi)=-\frac{1}{4}g_4^2 \sum\limits_{a,b}
\, [\Phi_a,\Phi_b]^2, \ee while the second term corresponds to an 
explicit $R$-symmetry-breaking potential, which breaks the ${\cal N}=4$ 
supersymmetry as well as the global $SU(4)$ symmetry. Let us mention that here and in the following we use the conventions of \cite{wipf}.

In the above expressions $\mu,\nu=0,1,2,3$ are four-dimensional spacetime 
indices and $D_\mu = \partial_\mu - i g [A_\mu,.]$ is the four-dimensional 
covariant derivative in the adjoint representation. The projection operators
 $L$ and $R$ are, as usual, defined as $L= \frac 12 (\one - \gamma_5)$ and $R= 
\frac 12 (\one + \gamma_5)$. The
$(\Delta_{L}^a)_{pq}$ and $(\Delta_{R}^a)_{pq}$ are the intertwiners of
the $\mathbf{4}\times\mathbf{4} 
\to \mathbf{6}$ and $\mathbf{\bar 4}\times\mathbf{\bar 4} \to \mathbf{6}$ 
respectively, namely they are Clebsch-Gordan coefficients that couple two 
$\mathbf{4}$s to a $\mathbf{6}$. The Yukawa interactions in (\ref{YM-action-4D}) 
are separately invariant
under the $SU(4)$, since the $R \chi_p$ transforms in the $\mathbf{4}$
and the $L\chi_p$ in the $\mathbf{\bar 4}$ of the $SU(4)$.

The action \eq{YM-action-4D} and the considerations below are
best understood through dimensional reduction,
starting from the ${\cal N}=1$ SYM theory in ten dimensions
with a ten-dimensional Majorana-Weyl spinor $\Psi$
and action
\be\label{tendaction}
S_{D=10} = -\frac 1{4g^2_{10}} \int d^{10} x\, \Tr F_{MN} F^{MN}\,\,
  + \,\,\frac 12 \int d^{10} x\, \Tr \bar\Psi i\Gamma^M  D_M \Psi ,
\ee
where
\be
 D_M = \partial_M - ig [A_M,.]
\ee and capital Latin letters denote ten-dimensional indices, i.e. $M= 0, .., 9$.
Considering a compactification of the form $M^4\times Y$, the scalars are obtained 
from the internal components of the higher-dimensional gauge field according to 
the splitting
\be
A_M = (A_\mu,\Phi_{3+a}), \quad a=1,\dots,6.
\ee
The ten-dimensional Clifford algebra, generated by
$\Gamma_M$, naturally separates into
a four-dimensional and a six-dimensional one as follows,
\bea
\Gamma_M&=&(\Gamma_{\mu},\Gamma_{3+a}), \nn\\ \Gamma_{\mu}&=&\gamma_
{\mu}\otimes\one_8, \nn\\ \Gamma_{3+a}& =& \gamma_5\otimes\Delta_a.
\eea
Here the $\gamma_{\mu}$ define the four-dimensional Clifford algebra and they are chosen 
to be purely imaginary,
corresponding to the Majorana representation in four dimensions (see Appendix A), 
while the $\Delta_a$ define the six-dimensional Euclidean Clifford
algebra and they are chosen to be real and antisymmetric.
Then it is straightforward to see that $\gamma_0 =
\gamma_0^\dagger= - \gamma_0^T$ and
$\gamma_i = -\gamma_i^\dagger=  \gamma_i^T$.
The ten-dimensional chirality operator is
\be
\Gamma^{(11)} = \gamma_5 \otimes\Gamma^{(Y)},
\ee
where the four- and six-dimensional chirality operators are defined as
\bea
\gamma_5 &=& -i\gamma_0 ... \gamma_3 = \gamma_5^\dagger
= - \gamma_5^T, \nn\\
\Gamma^{(Y)} &=& -i\Delta_1 ... \Delta_6 = (\Gamma^{(Y)})^\dagger
= -(\Gamma^{(Y)})^T.
\eea
Let us denote the ten-dimensional charge conjugation operator as
\be
\cC = C^{(4)} \otimes C^{(6)}, \ee
where $C^{(4)}$ is the four-dimensional charge
conjugation operator and $C^{(6)}=\one_8$ in our conventions. This operator 
satisfies, as usual, the relation
\be \cC\Gamma^M\cC^{-1}=-(\Gamma^M)^T. \ee Then
the Majorana-Weyl condition in ten dimensions is\footnote{Note that $T$ transposes only the spinor.}
\be
\Psi^C = \cC \bar\Psi^T
 {=} \Psi, \ee where
\bea\bar\Psi &=& \Psi^T \cC^T, \nn\\
\Psi^\dagger &=& \Psi^T \cC^T \gamma_0=\Psi^T.
\eea
Let us note that in the Majorana representation, where the $\gamma_\mu$
are imaginary, the four-dimensional charge conjugation operator is 
$C^{(4)} = -\gamma_0$. 

Performing a trivial dimensional reduction from ten to four dimensions, i.e. 
assuming that all fields do not depend on the internal coordinates, it is 
well-known that the Yang-Mills part of the ten-dimensional action leads to 
the bosonic part of the ${\cal N}=4$ SYM in four dimensions, as in
 (\ref{YM-action-4D}), with the potential term having the form (\ref{n4pot}). 
The couplings $g_4$ and $g_{10}$ are related through the volume $V$ of the 
internal six-dimensional torus as $g_4=\frac{g_{10}}{\sqrt{V}}$.

The reduction of the Dirac term is performed similarly. The Majorana-Weyl 
spinor $\Psi$
has the form
\bea
\Psi &=& \sum_{p=1}^4\,\Big( R\chi_p \otimes  \eta_p
 + L\chi_p \otimes  \,\eta_p^* \Big) , \nn\\
\bar\Psi &=& \sum_{p=1}^4\,\Big(\bar\chi_p L \otimes  \eta_p^\dagger
 + \bar\chi_p R \otimes  \,\eta_p^T \Big),
\label{Psi-4D}
\eea
where the $\chi_p$ are four-dimensional Majorana spinors
and the $\eta_p$ are the four complex eigenvectors of the $\Gamma^{(Y)}$ with eigenvalue
$+1$. Since the $\Gamma^{(Y)}$ is purely imaginary the $\eta_p^*$ have eigenvalue $-1$. 
Assuming that the spinor is independent of the extra-dimensional coordinates, 
the dimensional reduction of the Dirac term of the ten-dimensional action leads 
in four dimensions to the kinetic term for the spinor $\chi_p$ and the Yukawa 
couplings as they appear in (\ref{YM-action-4D}). In particular, the Yukawa 
couplings arise from the term
\bea
\Tr\, \bar\Psi  i \slashed D_{(6)}\Psi
= \Tr \,\bar\Psi  \,\Delta^{a}[\Phi_a,\Psi],
\label{6D-Yukawa}
\eea where $\slashed D_{(6)}$ denotes the Dirac operator on the internal space, 
which satisfies
\be
\{\slashed D_{(6)},\Gamma^{(Y)}\} =0
\label{D-6-chirality}
\ee and it will be related in the ensuing to the effective Dirac operator on the fuzzy
extra dimensions.

The above procedure, based on a trivial dimensional reduction, leads to the
 ${\cal N}=4$ SYM theory in four dimensions. However, in general, non-trivial
 dimensional reduction schemes \cite{Scherk:1979zr,Forgacs:1979zs,Witten:1985xb} are known to lead in four dimensions
 to actions preserving less supersymmetry or no supersymmetry at all.
 The choice of the reduction scheme is reflected in the four-dimensional vacuum. 
Correspondingly, in our case we add to the potential of the four-dimensional 
theory the term $V_{break}$, with the obvious requirement that it is 
renormalizable and moreover that it leads to vacua which can be interpreted as switching on fluxes in the internal extra-dimensional manifold. This is expected to be appropriate in order to obtain chiral zero modes.

\subsection{Type I vacuum and Dirac operator on fuzzy $S^2 \times S^2$}

\subsubsection{The type I vacuum}

We now assume that the renormalizable potential in the four-dimensional action
admits vacua corresponding to the product of two 
fuzzy spheres \cite{Madore:1991bw}, i.e.
\bea
\Phi_i^L &\equiv& \Phi_i \,\,\,\,\,\,=\,\, \a_L\, \la_i^{(N_L)}\otimes \one_{N_R}
\otimes\one_{n},\nn\\
\Phi_{i}^R &\equiv& \Phi_{3+i} \,\, = \,\,\a_R\, \one_{N_L}\otimes\la_{i}^{(N_R)}
\otimes\one_{n}, \qquad i=1,2,3,
\label{vacuum-typeI-S2S2}
\eea
where $\la_i^{(N)}$ denotes the generator of the
$N$-dimensional irreducible representation of $SU(2)$ and therefore
\bea
\, [\Phi_i^L,\Phi_j^L] &=& i \a_L\, \varepsilon_{ijk} \Phi_k^L,\nn\\ \Phi_i^L 
\Phi_i^L &=& \a_L^2 \frac{N_L^2-1}4,
\label{fuzzysphere}
\eea
and similarly for the $\Phi_i^R$. Moreover the two algebras commute with each other,
 i.e. \be  [\Phi_i^L,\Phi_j^R] = 0. \ee
This vacuum preserves the global $SO(3)_L \times SO(3)_R$ symmetry
up to gauge transformations:
\be
g\triangleright \Phi_i = U \Phi_i U^{-1},
\ee
where $g \triangleright$ denotes the action of some
$g \in SO(3)_L \times SO(3)_R$ in the vector indices of the $\Phi_i$
and $U\in U(N)$ is the corresponding intertwining gauge group action.

The vacuum \eq{vacuum-typeI-S2S2} can be obtained by choosing the potential $V(\Phi)$ 
to have the following form\footnote{This includes 
 purely soft deformations
from the $\cN=4$ potential, which are recovered
via $a_L \to 0, \, a_L b_L = const$. 
The present form emphasizes the case of true minima without flat
directions. In general, the potential breaks supersymmetry completely,
while some SUSY may be preserved for special choices of the parameters 
\cite{Andrews:2005cv}.}, 
\beq
V[\Phi] = a_L^2 (\Phi_i^L\Phi_i^L + b_L\one)^2
 + a_R^2 (\Phi_i^R\Phi_i^R + b_R\one)^2
 + \frac 1{g_L^2} F_{ij}^L F_{ij}^L + \frac 1{g_R^2} F_{ij}^R F_{ij}^R, \label{V-S2S2}
\eeq where
\bea
F_{ij}^L &=& [\Phi_i^L,\Phi_j^L] - i\varepsilon_{ijk} \Phi_k^L, \nn\\
F_{ij}^R &=& [\Phi_i^R,\Phi_j^R] - i\varepsilon_{ijk} \Phi_k^R,
\eea which will be interpreted as field strengths on the spontaneously generated fuzzy spheres.
The potential (\ref{V-S2S2}) breaks the global $SO(6)$ symmetry down to $SO(3)_L \times SO(3)_R$ and for suitable parameters $a_{L/R}, b_{L/R}, g_{L/R}$, its stable global minimum is indeed given by \eq{vacuum-typeI-S2S2} up to
$U(N)$ gauge transformations,
provided that
\be
N = N_L N_R\, n .
\label{typeI-condition}
\ee
Such a vacuum should be interpreted as a stack of $n$
fuzzy branes with geometry $S^2_L \times S^2_R$. In the present construction it breaks the
gauge group $SU(N)$ down to $SU(n)$.
The $\Phi_i^{L/R}$ are interpreted as
quantization of
the coordinate functions $x^i$ on $S^2_{L/R} \subset \R^3$ with
radius $R_{L/R}$.
More generally, there is a de-quantization map
\bea
Mat(N,\C) \, &\hookrightarrow& \cC(S^2) \nn\\
 \Phi_i    &\mapsto&   x_i
\eea
which extends to the spherical harmonics (defined as symmetric
traceless polynomials in $\Phi_i$ or, respectively, in $x_i$)
up to a cutoff. In this way, the matrices $Mat(N,\C)$
can be identified with functions on $S^2$, which is the basis of
the mechanism under consideration. This semi-classical limit, i.e. the limit $N\rightarrow\infty$,  
will be denoted as $\sim$ in the following.
In complete analogy to previous work
\cite{Aschieri:2006uw,Steinacker:2007ay}, it is not hard to see
that the model \eq{YM-action-4D}
can be interpreted in such a
vacuum as $SU(n)$ gauge theory on $S^2_L \times S^2_R$, via the identification
\bea
Mat(N_L N_R,\C) \, &\hookrightarrow& \cC(S^2_L \times S^2_R) \nn\\ 
 f(\Phi_a)    &\mapsto&   f(x_i^L,x_i^R) 
\eea
so that $Mat(N,\C)$ can be interpreted as $U(n)$-valued
functions on $S^2_L \times S^2_R$.
This is not surprising in view of the
origin of \eq{YM-action-4D} from dimensional reduction of
$D=10$ Yang-Mills theory.


\subsubsection{Operators on $S^2_L\times S^2_R$}

Having in mind the compactification on
$S^2_L\times S^2_R \subset \R^6$, we organize the internal
$SO(6)$ structure according to its subgroup
$SO(3)_L \times SO(3)_R$ and we adopt the notation
\be
\Delta_i^L = \Delta_i, \qquad \Delta_i^R = \Delta_{3+i},
\qquad i=1,2,3 .
\ee
Let us note that we have to work with the
six-dimensional Clifford algebra acting on $\C^8$,
which does not separate into a tangential and
transversal part. This is typical for fuzzy geometries.
In the following we shall rewrite the Dirac operator $\slashed D_{(6)}$
in terms of fuzzy Dirac operators on $S^2_L \times S^2_R$,
which allows to organize the fermionic Hilbert space
in terms of Kaluza-Klein modes.
Moreover, this allows to make explicit the role of the
would-be zero modes in the presence of fluxes which are
crucial in our context. However, the separation of
tangential and perpendicular quantities with respect to
$S^2_L \times S^2_R \subset \R^6$
is somewhat intricate and non-standard.

Let us consider the following $SO(3)_L \times SO(3)_R$
invariant operators on each sphere \cite{Behr:2005wp},
\bea
\chi_{L}\, &=&\, \frac i{2R_L}\, \Delta_i^L \{\Phi_i^L,.\}
\,\,\sim\,\, \frac i{R_L}\, \Delta^i_L x_i^L ,  \nn\\
\chi_{L,tang} &=& \Gamma^{(Y)}_L \chi_L, \nn\\
\Gamma^{(Y)}_{L} &=& \Delta_1 \Delta_2 \Delta_3,
\label{chi-L-def}
\eea 
where
\be
R_L = \a_L N_L
\ee
denotes the radius of $S^2_L$
and $\chi_{R}$, $\chi_{R,tang}$ and $\G^{(Y)}_R$ are defined similarly.
 These operators
are hermitian, i.e.
\bea
(\chi_{L/R})^\dagger&=&\chi_{L/R}, \nn\\
(\chi_{L/R,tang})^\dagger&=&\chi_{L/R,tang}, \nn\\
(\Gamma^{(Y)}_{L/R})^\dagger &=&\Gamma^{(Y)}_{L/R} ,
\eea and they satisfy the relations
\bea
\{\chi_{L/R}, \Gamma^{(Y)}\} \, &=&\, [\chi_L \chi_R,\Gamma^{(Y)}] = 0
\eea
and
\bea
\{\chi_L, \chi_R\} &=& 0  ,   \nn\\
\,[\chi_{L,tang}, \chi_{R,tang}] &=& 0, \nn\\
\chi_{L/R}^2 \sim & \one & \,\sim  \chi_{L/R,tang}^2
\eea
in a  $S^2_L \times S^2_R$ vacuum \eq{vacuum-typeI-S2S2}.
In order to understand their
meaning, let us consider e.g. $S^2_L$ as discussed in Appendix B.
On the north pole with $x_1\sim0, x_2\sim 0, x_3 \sim R_L$,
the tangential chirality operator is given by
$\chi_{L,tang}\sim i\Delta_1 \Delta_2$, while
the operator $\chi_{L} \sim i\Delta_3$ is perpendicular.
Therefore the $SU(2)_L \times SU(2)_R$-invariant operator
\be
\chi^\perp := i \chi_{L}\chi_{R},
\ee
which squares to one, $(\chi^\perp)^2 \sim 1$, corresponds to the chirality 
operator on the
two-dimensional space which is perpendicular to
$S^2_L\times S^2_R \subset \R^6$. In addition, the operator
\bea
\chi_{\rm tang} &:=& \Gamma^{(Y)}\chi^\perp
= - \chi_{L,tang} \chi_{R,tang}
\sim \Delta_1 \Delta_2\Delta_4 \Delta_5, \nn\\ (\chi_{\rm tang})^\dagger
 &=&\chi_{\rm tang}
\label{chi-tang-def}
\eea
is the tangential chirality operator on $S^2_L \times S^2_R$.

\subsubsection{The Dirac operator}

In order to understand 
the fuzzy Kaluza-Klein modes and especially the would-be zero modes,
it is important to understand the
relation of $\slashed D_{(6)}$ with the fuzzy Dirac operators
on $S^2_L$ and $S^2_R$. We note that
in the Majorana representation of the six-dimensional Clifford algebra, we have
\be
-i\Gamma^{(Y)}_L \D_i^L = \one_2 \otimes \gamma_L^i,
\ee
where $\gamma_L^i = U^{-1} (\sigma_i\otimes \one_2) U$
is essentially a double-degenerate
representation of the three-dimensional Clifford algebra.
This allows to write
\be
\Delta^i_{L} [\Phi_i^{L},\Psi]
+ i\a_{L}\Gamma^{(Y)}_{L} = i\Gamma^{(Y)}_L \slashed D_{S^2_{L}},
\label{D-S2-6D}
\ee
where $\slashed D_{S^2_{L}}$ is the standard Dirac operator on
the fuzzy $S^2_L$ (see Appendix B for a short review).
Note that one usually works with two-component
spinors on the fuzzy sphere, where the tangential chirality operator
is given by $\sigma_1 \sigma_2 =i \sigma_3$. Here
$\Delta_3$ is independent of $\Delta_1 \Delta_2$ and therefore
$\chi_{L,tang}$
is the proper tangential chirality operator
on the $S^2_L$, rather than $\chi_{L}$. In particular, the would-be zero modes
$\Psi^{1,2}_{(m)}$ on the $S^2_L$ with magnetic flux, discussed in Appendix B, are
eigenvectors of $\chi_{L,tang}$ and not
eigenvectors of $\chi_L$\footnote{Note that the
generator of $SU(2)_L$ on the spinors
is given by
$[\Delta^i_L,\Delta^j_L] = - 2 i \varepsilon^{ijk} \one_2 \otimes
\gamma_L^k$.}.
We thus obtain the relation of $\slashed D_{(6)}$
with a ``tangential'' Dirac operator on
$S^2 \times S^2 \subset \R^6$:
\be
\slashed D_{(6)} = \slashed D_{S^2\times S^2} \,\,
 - \,\, \a_L \Gamma^{(Y)}_L - \a_R \Gamma^{(Y)}_R,  \ee where \be
\slashed D_{S^2\times S^2}
=    \Gamma^{(Y)}_L\slashed D_{S^2_L}
 + \Gamma^{(Y)}_R\slashed D_{S^2_R}.
\label{6Ddirac-mod}
\ee
Then the Yukawa term becomes
\be \cS_{yuk}= \int \obar \Psi i \slashed D_{S^2\times S^2} \Psi+\cS_{shift}, \ee
where the shift action
\be
\cS_{shift}= \int i Tr\,\obar\Psi\gamma_5(\a_L \Gamma^{(Y)}_L + \a_R\Gamma^{(Y)}_R)
\Psi
\label{S-shift}
\ee
is recognized as curvature effect.
One can show that $\slashed D_{S^2\times S^2}$ reduces
in the semi-classical limit to the Dirac operator on $S^2_L \times S^2_R$
in the above background geometry \eq{vacuum-typeI-S2S2}
\cite{Behr:2005wp}.
Note also that
\be
\cS_{shift,L}^* = \int (i Tr\,\Psi^\dagger\gamma_0\g_5 \alpha_L\Gamma^{(Y)}_L\Psi)^\dagger
= \int i Tr\,\Psi^\dagger\gamma_0 \g_5\alpha_L\Gamma^{(Y)}_L\Psi
=  \cS_{shift,L}
\ee and similarly for the $\cS_{shift,R}$, as expected since the original action is hermitian.

\section{Magnetic fluxes and chirality}
\label{sec:typeII-vac}

\subsection{The type II vacuum and the zero-modes}

In order to obtain massless fermions, it is necessary to add
magnetic fluxes $m_L, m_R$ on the two spheres.
As explained in \cite{Steinacker:2007ay}, this is realized in a slightly modified class
of vacua, called ``type II vacua''.
In the present case such a vacuum has the form
\bea
\Phi_i
&=& \left(\begin{array}{cc}\a_1\, \la_i^{(N_L^1)}\otimes \one_{N_R^1}
\otimes\one_{n_1} & 0 \\ 0
& \a_2\,\la_i^{(N_L^2)}\otimes \one_{N_R^2}\otimes\one_{n_2}
             \end{array}\right),  \nn\\
\Phi_{3+i}
&=& \left(\begin{array}{cc}\a_3\, \one_{N_L^1}\otimes\la_{i}^{(N_R^1)}
\otimes\one_{n_1} & 0 \\ 0
& \a_4\,\one_{N_L^2}\otimes\la_{i}^{(N_R^2)}\otimes\one_{n_2}
             \end{array}\right),\quad i=1,2,3.  \nn\\
\label{vacuum-mod2_s2s2}
\eea
The commutant of these generators, i.e. the
unbroken gauge group, is
$SU(n_1) \times SU(n_2) \times U(1)_Q$, where the $U(1)_Q$
has generator
\be
Q = \left(\begin{array}{cc} \frac 1{N_R^1 N_L^1 n_1}\one & 0 \\ 0
& -\frac 1{N_R^2 N_L^2 n_2}\one
             \end{array}\right).
\ee
This vacuum corresponds to a splitting
\be
 N = n_1 N_L^1 N_R^1 + n_2 N_L^2 N_R^2
\ee and, respectively, \be \cH = \C^{n_1 N_L^1 N_R^1} \oplus \C^{n_2 N_L^2 N_R^2} \ee
for the Hilbert space,
which is more generic than \eq{typeI-condition}. It
determines a splitting of the
fermionic wavefunction
\be
\Psi = \left(\begin{array}{cc} \Psi^{1,1} & \Psi^{1,2} \\
                              \Psi^{2,1} & \Psi^{2,2}
\end{array}\right),
\ee
where $\Psi^{1,2}$ transforms in the bifundamental representation $(n_1)\otimes 
(\obar n_2)$ of the
$SU(n_1) \times SU(n_2)$ and
$\Psi^{2,1}$  in the $(\obar n_1)\otimes (n_2)$.
The Majorana condition $\Psi^+\equiv\Psi^{\dagger T} = \Psi$ implies
\bea
(\Psi^{1,1})^{+} =\Psi^{1,1},\quad
(\Psi^{2,2})^{+} =\Psi^{2,2},\quad
(\Psi^{1,2})^{+} =\Psi^{2,1} .
\label{MW-block}
\eea
The interpretation of this vacuum is as a stack of
$n_1$ fuzzy branes
and a stack of $n_2$ fuzzy branes\footnote{This is quite 
reminiscent of standard constructions
in the context of string theory, see e.g. 
\cite{Blumenhagen:2006ci}.
However, there are several differences to the situation in
string theory. Notably, the branes do not live in a ten-dimensional
space; even though
$M^4 \times S_L^2\times S_R^2 \subset M^4 \times \R^6$,
the two ``missing dimensions'' in $\R^6$
have no physical meaning here and carry
no physical degrees of freedom.} with geometry $S^2_L \times S^2_R$.
However, these fuzzy spheres carry
magnetic flux under the unbroken $U(1)_Q$
given by \cite{Steinacker:2003sd}
\be
m_L= N_L^1-N_L^2, \qquad m_R = N_R^1-N_R^2,
\ee
on $S^2_L$ and $S^2_R$ respectively.
Since the fermions $\Psi$ transform in
the adjoint representation, the diagonal components $\Psi^{1,1}$ and
$\Psi^{2,2}$ are unaffected, but the
off-diagonal components $\Psi^{1,2}$ and
$\Psi^{2,1}$ feel this magnetic flux and develop chiral zero
modes according to the index theorem. This can also be seen very
explicitly in the fuzzy case \cite{Steinacker:2007ay}, see Appendix B.
For example, a flux $m_L> 0$ on  $S^2_L$ implies
that there are (would-be) zero modes $\Psi^{1,2}_{(m_L)}$
for $\slashed D_{S^2_{L}}$ with $\chi_{L,tang} = +1$, and $\Psi^{2,1}_{(m_R)}$
with $\chi_{L,tang} = -1$.

To be specific, assume that  $m_L>0$ and $m_R>0$.
As explained in Appendix B, then 
there exist  (``would-be'', approximate) 
zero modes $\Psi^{1,2}_{(m_L,m_R)}$
of both $\slashed D_{S^2_{L}}$
and $\slashed D_{S^2_{R}}$ and therefore of 
$\slashed D_{S^2\times S^2}$,
with definite chirality\footnote{To simplify the notation we assume
that the operators $\chi, \slashed D_{S^2}$ 
are defined appropriately such that
these relations hold exactly, see Appendix B. Otherwise 
the stated eigenvalues of $\chi$ and $\slashed D_{S^2}$
are approximate up to $O(\frac 1N)$ corrections. 
Since we are mainly interested
in the structure of the would-be zero modes, we do not keep 
track of these $O(\frac 1N)$ corrections here.}
\bea
\chi_{L,tang} \Psi^{1,2}_{(m_L,m_R)} &=&   \Psi^{1,2}_{(m_L,m_R)}
= \chi_{R,tang} \Psi^{1,2}_{(m_L,m_R)}, \nn\\
\chi_{tang} \Psi^{1,2}_{(m_L,m_R)}  &=&  \Psi^{1,2}_{(m_L,m_R)}.
\label{zeromodes-full-12}
\eea
There are also the ``conjugate'' zero modes
$\Psi^{2,1}_{(m_L,m_R)}$, which satisfy
\bea
\chi_{L,tang} \Psi^{2,1}_{(m_L,m_R)} &=& - \Psi^{2,1}_{(m_L,m_R)}
= \chi_{R,tang} \Psi^{2,1}_{(m_L,m_R)},  \nn\\
\chi_{tang} \Psi^{2,1}_{(m_L,m_R)}  &=&   \Psi^{2,1}_{(m_L,m_R)}.
\label{zeromodes-full-21}
\eea
Thus, in general, we have
\be
\chi_{tang} \Psi_{(m_L,m_R)} =  (-1)^{m_L+m_R}\,  \Psi_{(m_L,m_R)}.
\label{chi-tang-zero}
\ee
All the other Kaluza-Klein modes have both chiralities and acquire a mass due to $\slashed D_{S^2\times S^2}$.

Motivated by the properties of the zero modes which are encoded in 
(\ref{zeromodes-full-12}) and (\ref{zeromodes-full-21}) let us now define 
the following operators,
\bea
\Pi_L \Psi := \gamma_5\chi_{L,tang}\Psi,  \nn\\
\Pi_R \Psi := \gamma_5\chi_{R,tang}\Psi,\label{chiral-map}
\eea
which satisfy 
\be
 \Pi_L^2 \sim \one \sim \Pi_R^2.
\ee
They are clearly compatible with the ten-dimensional
Weyl condition and also with the ten-dimensional Majorana condition
$\Psi^\dagger = \Psi^T$, since
\bea
(\Pi_L \Psi)^\dagger = (\Pi_L \Psi)^T, \nn\\
(\Pi_R \Psi)^\dagger = (\Pi_R \Psi)^T.
\eea
Consequently they are well-defined and as we shall exhibit in the following 
they will select the chiral sectors of our model.
In order to understand the qualitative structure of the zero modes, in particular their chirality from the four-dimensional point of view, it is enough to consider the semi-classical limit.
On the north pole we have, for the
adapted representation given in the Appendix A,
\bea
\chi_{L,tang} &\sim& i\D_1 \D_2 =  \one\otimes \sigma^3 \otimes \sigma^2,\label{chitangL}\\
\chi_{R,tang} &\sim& i\D_4 \D_5 =  \one\otimes \sigma^2 \otimes \sigma^3,\label{chitangR}\\
\chi_{tang} &\sim&  \one\otimes \sigma^1 \otimes \sigma^1, \\
\chi^\perp &\sim& \sigma_2\otimes \sigma^1 \otimes \sigma^1.
\eea
Then, the unique solution of  \eq{zeromodes-full-12} has the form
\be
\Psi^{1,2}_{(m_L,m_R)} \sim\left(\barr{l}\rho^{1,2}\\ \eta^{1,2}\earr\right)
\otimes \left(\left(\barr{l}1\\1\earr\right)\otimes\left(\barr{l}1\\1\earr\right)
 -  i\left(\barr{l}1\\-1\earr\right)\otimes\left(\barr{l}1\\-1\earr\right)\right),
\label{zeromodes-spinor-12}
\ee
where $\rho^{1,2}, \eta^{1,2}$ are four-dimensional Dirac spinors. Similarly, 
the unique solution of  \eq{zeromodes-full-21} has the form
\be
\Psi^{2,1}_{(m_L,m_R)} \sim\left(\barr{l}\rho^{2,1}\\ \eta^{2,1}\earr\right)
\otimes \left(\left(\barr{l}1\\1\earr\right)\otimes\left(\barr{l}1\\1\earr\right)
 +  i\left(\barr{l}1\\-1\earr\right)\otimes\left(\barr{l}1\\-1\earr\right)\right).
\label{zeromodes-spinor-21}
\ee
The Weyl condition $\Gamma^{(11)}\Psi = \Psi$ implies
\bea
\gamma_5\Psi  &=& \Gamma^{(Y)} \Psi
 \,\, =\,\,  - (\sigma_2\otimes\one\otimes\one) \Psi,  \nn\\
i\eta^{1,2} &=& \gamma_5\rho^{1,2},\nn\\
i\eta^{2,1} &=& \gamma_5\rho^{2,1},
\label{Weyl-explicit}
\eea
so that the would-be zero mode reduces essentially to 
\be
\Psi_{(m)}^{1,2} \sim \left(\barr{l}\rho^{1,2}\\
  -i \gamma_5 \rho^{1,2}\earr\right), \qquad
\obar{\Psi_{(m)}^{1,2}} \sim (\obar{\rho^{1,2}},
  -i \obar{\rho^{1,2}}\gamma_5 ),
\ee
dropping the remaining tensor factors in \eq{zeromodes-spinor-12}. 
The Majorana condition, $\Psi^+:=\Psi^{\dag T}=\Psi$, in the present
representation implies
\be
\rho^{1,2} = (\rho^{2,1})^+
\label{CC-explicit}
\ee
and it relates the upper-diagonal and lower-diagonal components.
This amounts to a single four-dimensional Dirac spinor $\rho^{1,2}$ and
the model is non-chiral.

\subsection{Chiral low-energy theory.}

In general, a model is chiral if  all left-handed Weyl
spinors $\psi_\a$ (including those obtained by
conjugation of right-handed ones)
live in the same complex representation of the gauge group
and therefore a mass term $\sim \psi_\a \chi_\b \varepsilon^{\a\b}$ cannot be written
 down. 
However, here we work with Majorana spinors, which necessarily
involve both chiralities, and one has to be somewhat careful.

Let us consider again the would-be zero modes.
The cleanest way to understand their structure is to look at the upper
(respectively lower) triangular matrices $\Psi^{1,2}$ (respectively $\Psi^{2,1}$),
which are in inequivalent (conjugate) complex representations
of the unbroken gauge group $SU(n_1) \times SU(n_2)$.
They are related by the Majorana condition
$\Psi^{1,2}=(\Psi^{2,1})^+$.
Therefore we certainly cannot
expect to have $\Psi^{2,1}=0$. 
What is actually needed is that $\gamma_5|_{\Psi^{1,2}} = +1$
and $\gamma_5|_{\Psi^{2,1}} = -1$, so that the upper-triangular
matrices have chirality $+1$, and the lower-triangular matrices
are their charge conjugate modes; this is then a chiral theory.

Now let us show how to realize this.
We recall that we have two fuzzy spheres with fluxes and we have
assumed already that $m_L>0$ and $m_R>0$.
Then, the relations (\ref{zeromodes-full-12}) and (\ref{zeromodes-full-21}) can be
written as
\bea
{\chi_{L,tang}}|_{\Psi^{1,2}} &=& \chi_{R,tang}|_{\Psi^{1,2}} = +1 ,  \nn\\
{\chi_{L,tang}}|_{\Psi^{2,1}} &=& \chi_{R,tang}|_{\Psi^{2,1}} = -1.
\eea
It follows from (\ref{zeromodes-full-12}) and (\ref{zeromodes-full-21}) that the operators $\Pi_L$ and $\Pi_R$, defined in (\ref{chiral-map}), actually coincide on the space of zero modes.
Hence the full fermionic Hilbert space can be separated in two sectors as follows,
\be
\cH_+=\{\Psi; \quad \Pi_L\Psi = \Psi\} \qquad \mbox{and}\qquad
\cH_-=\{\Psi; \quad \Pi_L\Psi = -\Psi\}. \label{chiralsectors}
\ee
Then it is clear that $\Psi^{1,2}$ and $\Psi^{2,1}$ have opposite four-dimensional 
chirality in each sector, which is the desired result as it was explained above. 
Therefore we end up with two exactly chiral mirror sectors, which are separated according to (\ref{chiralsectors}). 
One can show that in a type II vacuum the fermion spectrum of the standard model can be accomodated in these sectors
\cite{Grosse2009}. Then, if the electroweak symmetry
breaking occurs in one sector at a scale which is much
higher than in the other sector, such a model might turn out to be
 realistic, which remains to be studied.

However, in order to be able to proceed further with the electroweak symmetry breaking,
the two chiral sectors that we have presented above have to decouple. Concerning 
the kinetic part of the Yukawa couplings 
it is straightforward to show 
that \be Tr\overline{\Pi_L\Psi}\g_5i\slashed D_{S^2\times S^2} \Psi'=
Tr\overline{\Psi}\g_5i\slashed D_{S^2\times S^2} \Pi_L\Psi'. \ee
Therefore the Yukawa term due to $\slashed D_{S^2\times S^2}$ couples only modes with
the same eigenvalue of $\Pi_L$. This is as expected for a kinetic term for spinors on
$S^2\times S^2$.

Finally, one might worry that the would-be zero modes are not massless due to the
 presence of the shift action (\ref{S-shift}) in the model and therefore that the chiral sectors 
are not exactly separate.
Indeed, the shift action couples
opposite eigenvalues of $\Pi_L$, since 
\be Tr\overline{\Pi_L\Psi}\g_5i\G_L^{(Y)}\Psi'=
-Tr\overline{\Psi}\g_5i\G_L^{(Y)}\Pi_L\Psi'. \ee 
Therefore, in order to make the would-be
zero modes massless, we have to modify the model
by adding a
mass term which breaks $SO(6)\cong SU(4) \to SU(2)_L \times SU(2)_R$, in which case 
the correct action for a chiral model is given by
\bea
S_{chiral}[\Psi] &=& S - S_{shift}  \quad
= \quad \int \obar \Psi i\slashed D_{(6)} \Psi
- \obar \Psi i\g_5(\alpha_L\Gamma^{(Y)}_L + \alpha_R\Gamma^{(Y)}_R) \Psi \nn\\
&=&  \int \obar \Psi i \slashed D_{S^2\times S^2} \Psi.
\eea
This action is singled out by the fact that the would-be zero modes are
exactly massless at tree level and therefore the model enjoys
unbroken $U(1)_{axial}$ symmetry generated by
$\Pi_L$. This means that quantum corrections
will not induce any mass terms since they are protected by chiral
symmetry. The two chiral sectors $\cH_+$ and $\cH_-$ are now 
exactly separated and one can proceed to study
the electroweak symmetry breaking in each one.  This will be 
pursued elsewhere. 

\section{Discussion and conclusions}

Generalizing previous work 
\cite{Aschieri:2006uw,Steinacker:2007ay,Aschieri:2003vy,Andrews:2005cv},
we have shown how an effective extra-dimensional space
with the geometry of fuzzy $S^2\times S^2$ can arise within
a renormalizable four-dimensional $SU(N)$
gauge theory with the matter content of $\cN=4$ SYM
theory.  The underlying mechanism is simply the standard Higgs effect and
spontaneous symmetry breaking of the theory. 
The model behaves as an eight-dimensional
 Yang-Mills theory on $M^4 \times S^2_N \times S^2_N$, 
for energies below some cutoff given by the fuzzy nature 
of the extra-dimensional spheres. 
 It represents a particularly
simple yet rich realization of the idea of deconstructing dimensions
\cite{Arkani-Hamed:2001ca}, taking advantage of 
results from noncommutative field theory.
This allows to consider ideas of compactification and 
dimensional reduction within a renormalizable framework.
We focus on the effects of fluxes on these fuzzy spheres
and on the corresponding fermionic zero modes, i.e. the low-energy
sector of the model.

Even though the fluxes on $S^2\times S^2$ lead indeed
to the expected zero modes, the model nevertheless
turns out to be non-chiral a priori.
More precisely, we find essentially mirror models, where 
two chiral sectors arise with opposite chirality. 
This means that each would-be zero mode from $\Psi^{1,2}$ 
has a mirror partner from $\Psi^{2,1}$, with opposite
chirality and gauge quantum numbers. 
The reason for this is that the fuzzy geometry 
is four-dimensional but in some sense embedded in six extra dimensions.
The missing two (``shadow'') dimensions are reflected 
in extra components of the spinors, which do not see the 
flux respectively the chirality on 
$S^2\times S^2$. This is a crucial difference of our model comparing 
with models based on commutative extra dimensions, 
where chiral Lagrangians are easier to obtain 
\cite{Chapline:1982wy,Kapetanakis:1992hf}\footnote{see however 
\cite{Barnes:1986ea}.}.
Thus we arrive essentially at a picture of mirror fermions 
discussed e.g. in \cite{Maalampi:1988va} from a 
phenomenological point of  view.
While this may still be physically interesting since the ``mirror fermions'' 
may have larger mass than the ones we see at low energies,
it would be desirable to find a chiral version 
with similar features. 
Thus the present work can be seen as a step in the 
direction of realistic models in this framework, suggesting
the need of additional geometrical structures in the extra
dimensions. There are indeed many possible 
directions for generalizations,
and we hope to report on progress in this direction soon.

\paragraph{Acknowledgments}

We are grateful for discussions with 
H. Grosse and F. Lizzi. 
This work is supported by the NTUA programme for basic research PEVE 2008 and the European Union's RTN programme under contract MRTN-CT-2006-035505.
The work of H.S. is supported in part by the FWF project
P20017 and in part by the FWF project P21610.   

\vspace{25pt}

\noindent {\bf\Large Appendices}

\renewcommand{\theequation}{A-\arabic{equation}}
  \setcounter{equation}{0}  

\section{Appendix A: Clifford algebra
and $SO(3)_L \times SO(3)_R$}

In this appendix we collect our conventions on the representations of the Clifford 
algebras we have used.

The gamma matrices $\gamma_{\mu}$ define the four-dimensional Clifford algebra and 
they are chosen to be purely imaginary, corresponding to the Majorana representation 
in four dimensions. In our conventions this representation is
explicitly given by the matrices
\bea \g_0&=&\s_0\otimes\s_2,\nn\\
       \g_1&=&i\s_0\otimes\s_3,\nn\\
       \g_2&=&i\s_1\otimes\s_1,\nn\\
       \g_3&=&i\s_3\otimes\s_1, \eea
where $\s_0:=\one_2$ is the identity matrix.

Moreover, we give here the explicit Majorana representation of the
six-dimensional Clifford algebra, which is known to exist in six Euclidean dimensions. This is naturally adapted to
$SO(3)_L \times SO(3)_R \subset SO(6)$, and closely related to other
constructions, see for example the ref. \cite{Behr:2005wp}. First we consider the
 matrices
\be \barr{llll}
&\gamma^{1}_{L}=\sigma ^{1}\otimes \sigma ^{2},
&\gamma^{2}_{L}=\sigma ^{2}\otimes \one,
&\gamma^{3}_{L}=\sigma^{3}\otimes \sigma ^{2}, \\
&\gamma^{1}_{R}=\sigma ^{2}\otimes \sigma ^{1},
&\gamma^{2}_{R}=-\sigma ^{2}\otimes \sigma ^{3},
&\gamma^{3}_{R}= \one\otimes \sigma ^{2},
\earr
\label{gamma-so4}
\ee
which are antisymmetric and purely imaginary,
hence hermitian, and they satisfy
\begin{eqnarray*}
\gamma _{L}^{i}\gamma _{L}^{j} & = & \delta ^{ij}+i\epsilon ^{ij}_{k}\gamma _{L}^{k},\\
\gamma _{R}^{i}\gamma ^{j}_{R} & = & \delta ^{ij}+i\epsilon ^{ij}_{k}\gamma ^{k}_{R},\\
{}[\gamma ^{i}_{L},\gamma ^{j}_{R}] & = & 0.
\end{eqnarray*}
Then the following matrices define a representation of the $SO(6)$ Clifford
algebra
\be
\D_i = i \sigma_1\otimes \gamma_{L}^{i},
\qquad \D_{3+i} = i \sigma_3\otimes \gamma_{R}^{i},
\label{6D-Majorana}
\ee
satisfying the desired relation
\be
\{\D^{\mu },\D^{\nu }\} = - 2\delta^{\mu \nu }.
\label{clifford-so6}
\ee
They are manifestly anti-symmetric and real, hence they furnish a Majorana representation.
The left and right chiral projections are given by
\bea
\Gamma_L^{(Y)} &=& \D_1\D_2\D_3 =  \sigma_1 \otimes \one,\nn\\
 \Gamma_R^{(Y)} &=& \D_4\D_5\D_6 =  \sigma_3 \otimes \one
\label{gamma-3-rep}
\eea
and the six-dimensional chirality operator is
\be
\Gamma^{(Y)} = -i\Gamma_L^{(Y)}\Gamma_R^{(Y)}
= -\sigma_2 \otimes \one.
\ee

\renewcommand{\theequation}{B-\arabic{equation}}
  \setcounter{equation}{0}  

\section{Appendix B: Dirac operator and eigenmodes on fuzzy $S^2$}

Assume that  $\phi_i$ satisfies the
relations of the fuzzy sphere 
\be
[\phi_i,\phi_j] = i \a\varepsilon_{ijk} \phi_k, \qquad
\phi_i \phi_i = \a^2\frac{N^2-1}{4} ,
\ee
and consider the Dirac operator on $S^2_N$ defined through
(cf. \cite{Grosse:1994ed,Balachandran:2005ew,Steinacker:2007ay})
\be
\slashed D_{S^2} \psi = \sigma^i [\phi_i,\psi] + \a\psi.
\label{fuzzy-Dirac-S2}
\ee
Since $\slashed D_{S^2}$ commutes with the $SU(2)$ group of rotations,
the eigenmodes of  $\slashed D_{S^2}$
are obtained by decomposing the spinors into irreducible representations of $SU(2)$,
\bea
\psi &\in&  (2)  \otimes (N) \otimes (N)
 = (2)\tens ((1)\oplus(3) \oplus ... \oplus (2N-1))\nn\\
&=& ((2) \oplus (4) \oplus ...  \oplus(2N))
 \, \oplus\,(\qquad \,\,\,(2) \oplus ... \oplus (2N-2))\nn\\
&=:& (\qquad \psi_{+,(n)}  \qquad \qquad \quad\,\,\,\,
        \oplus \qquad \psi_{-,(n)}) .
\label{spinor-decomp}
\eea
This decomposition defines the spinor harmonics $\psi_{\pm,(n)}$ which live in
the $n$-dimensional representation of $SU(2)$ denoted by $(n)$ for
$n=2,4,...,2N$, excluding $\psi_{-,(2N)}$. The eigenvalue
of $\slashed D_{S^2}$ acting on these states can be determined easily
using some $SU(2)$ algebra, 
\be
\slashed D_{S^2}\psi_{\pm,(n)}
= E_{\delta = \pm,(n)}\,\psi_{\pm,(n)} ,
\label{D-spectrum}
\ee
where
\be
E_{\delta = \pm,(n)}\, =\, \frac{\a}2\, \left\{\begin{array}{rl}
n, & \delta = 1,\qquad n=2,4,..., 2N \\
-n, & \delta = -1,\quad\, n=2,4,..., 2N-2
\end{array}\right. .
\label{D-spectrum-explicit}
\ee
We note that with the
exception of $\psi_{+,(2N)}$, all eigenstates come in pairs
$(\psi_{+,(n)},\psi_{-,(n)})$ for $n=2,4,..., 2N-2$, which have opposite
eigenvalues $\pm \frac{\a}2 n$ of $\slashed D_{S^2}$. They are
interchanged through the fuzzy 
chirality operator $\chi = \frac 1{2R} \sigma^i\{\phi_i,.\}$,
\be
\chi \left(\begin{array}{c} \psi_{+,(n)} \\ \psi_{-,(n)}
\end{array}\right) =
c \left(\begin{array}{cc} 0 & 1 \\
            1 & 0 \end{array}\right)
\left(\begin{array}{c} \psi_{+,(n)} \\ \psi_{-,(n)}
\end{array}\right),
\label{chi-exchange}
\ee
for some $c \approx 1$,
by virtue of the anticommutativity relation
\be
\slashed D_{S^2}\chi + \chi \slashed D_{S^2} =0 .
\label{D-chi-anticomm-gen}
\ee
 Moreover, the chirality operator for the top mode
 vanishes, $\chi \psi_{+,(2N)} =0$.

Let us now consider a type II vacuum 
analogous to \eq{vacuum-mod2_s2s2} in section \ref{sec:typeII-vac},
but for a simple fuzzy sphere \cite{Steinacker:2007ay}:
\be
\phi_i
= \left(\begin{array}{cc} \a_1\la_i^{(N_1)} & 0 \\ 0
& \a_2 \la_i^{(N_2)}
             \end{array}\right),
\label{vacuum-mod2_s2}
\ee
with $m = N_1-N_2$.
This can be interpreted as a two $U(1)$ gauge fields 
on $S^2$ which differ by a magnetic flux $m$ 
\cite{Steinacker:2003sd}.
Then there are $m$ ``would-be'' zero-modes 
as expected by the index theorem, 
which have a simple group-theoretical
realization. The point is that the
off-diagonal blocks are rectangular matrices, e.g.
$\psi^{1,2} \in Mat(N_1 \times N_2)$.
To be specific assume that $m>0$.  We
decompose this module, tensored with the spinorial $(2)$, 
into irreducible representations
 of the rotation group $SU(2)$,
\bea
(2) \otimes Mat(N_1 \times N_2) &\cong& (2) \otimes (N_1) \otimes
(N_2) \nn\\
 &\cong& (|N_1-N_2|) \oplus 2 \times (|N_1-N_2|+3)  \nn\\
&& \qquad \quad\,\, \oplus ...  \oplus 2 \times (N_1+N_2 -2)
\oplus (N_1+N_2) .
\label{zeromodes-decomp}
\eea
Note that there is indeed a single irreducible representation $(m)$,
which corresponds to a (would-be) zero mode 
that we denote as $\Psi^{1,2}_{(m)}$.
All the other irreducible representations come in pairs with opposite
chirality\footnote{Except for the top modes, where $\chi = 0$.}.
Together with $SU(2)$ invariance,
it follows that $\Psi^{1,2}_{(m)}$ is an
eigenstate of both the fuzzy Dirac operator $\slashed D_{S^2}$
and  the  fuzzy chirality operator $\chi$. 

In general, these would-be zero modes will be
only approximate zero modes, 
\bea
\slashed D_{S^2} \Psi^{1,2}_{(m)} &=& E_0\, \Psi^{1,2}_{(m)},
\qquad E_0 = O(\frac 1N),    \nn\\
\chi \Psi^{1,2}_{(m)} &=& c \Psi^{1,2}_{(m)}, \qquad \quad c \approx \pm 1
\eea
assuming $\a_i = 1+ O(\frac 1N)$.
One can either leave things in this approximate form, or 
cast it in a precise form assuming that 
$\a_1 N_1=\a_2 N_2 =: R$.
In that case, it is useful to consider
$\phi_0 = \frac 12\left(\begin{array}{cc}  \a_1 \one & 0 \\ 0
& \a_2 \one \end{array}\right)$ in the vacuum \eq{vacuum-mod2_s2},
which satisfies 
\be
\Phi = \phi_i\sigma^i + \phi_0 \sigma_0,  \qquad \Phi^2 = \frac{R^2}4 .
\ee
Then the following refined Dirac resp. chirality operators
\bea
\slashed D_{S^2} \psi &=& \sigma^i [\phi_i,\psi] + \{\phi_0,\psi\},  \nn\\
\chi \psi &=& \frac 1R\, (\sigma^i [\phi_i,\psi] + [\phi_0,\psi]  ) , 
\eea
satisfy 
\be
\{\slashed D_{S^2},\chi\} = 0
\ee
exactly. This implies that
$\Psi^{1,2}_{(m)}$ is an exact zero mode either of 
$\slashed D_{S^2}$ or of $\chi$.
It is easy to see (cf. below) that the latter does not hold, therefore
\be
\slashed D_{S^2} \Psi^{1,2}_{(m)} =0 .
\ee
We note that such zero modes on a sphere
only occur in the presence of magnetic flux. 
The eigenvalue of $\chi^2$ can be computed as follows:
we have 
$2\Phi \Psi^{1,2}_{-,(m)} = (R\chi + \slashed
D_{S^2})\Psi^{1,2}_{-,(m)} = R\chi \Psi^{1,2}_{-,(m)}$ on the zero modes. 
Using $4\Phi^2 = R^2$ it follows that 
\be
\chi^2 \Psi^{1,2}_{-,(m)} = \Psi^{1,2}_{-,(m)}.
\ee
The sign of $\chi$ on $\Psi^{1,2}_{(m)}$ is
easy to obtain.
Note that \eq{zeromodes-decomp} involves the ``anti-parallel''
(respectively ``parallel'') tensor product
of $(2) \otimes (N_1)$, and the ``parallel''
(respectively ``anti-parallel'') tensor product
of $(2) \otimes (N_2)$.
Therefore $\chi \sim \sigma^i x_i \approx {\rm sign}(m)$
when it acts on  $\Psi^{1,2}_{(m)}$  and similarly
$\chi\approx - {\rm sign}(m)$ when it acts on $\Psi^{2,1}_{(m)}$.
It follows that
\bea
\chi\Psi^{1,2}_{-,(m)} &=& + \Psi^{1,2}_{-,(m)},  \nn\\
\chi\Psi^{2,1}_{-,(m)} &=& - \Psi^{2,1}_{-,(m)} .
\label{zero-modes-chiral}
\eea
The inclusion of the second sphere is straightforward.
There are $m_L m_R$ (would-be) zero modes of 
$\slashed D_{S^2\times S^2}$ in the presence of 
a flux $m_L \neq 0$ on $S^2_L$ and a flux 
$m_R\neq 0$ on $S^2_R$,
given simply by the product of the above zero modes.

\end{document}